\documentclass[12pt,twoside]{article}

\evensidemargin=\oddsidemargin
\def\Title#1#2#3{%
    \baselineskip=18pt
    \begin{center}
          {\large\bf{#1} \\ }
          \bigskip\bigskip
          {#2} \\
          {#3} \\
    \end{center}}
\long\def\Abstract#1{%
         \bigskip
         \parbox{0.93\textwidth}{%
                 \begin{center}
                       {\bf Abstract} \\
                 \end{center}
                 \medskip{\baselineskip=14pt #1}
                 \vss}
         \bigskip}

\makeatletter
\renewcommand{\section}%
 {\@startsection{section}{1}{0pt}%
  {-3.25ex plus -1ex minus -.2ex}{1.5ex plus .2ex}%
  {\vspace*{5mm}\raggedright\large\bf }}
\renewcommand{\thesection}{\arabic{section}.}
\@addtoreset{equation}{section}
\renewcommand{\@eqnnum}{(\thesection\theequation)}
\renewcommand{\p@equation}{\thesection}
\makeatother

\begin{document}

\Title{On canonical transformations of gravitational variables\\
in extended phase space}%
{T. P. Shestakova}%
{Department of Theoretical and Computational Physics,
Southern Federal University,\\
Sorge St. 5, Rostov-on-Don 344090, Russia \\
E-mail: {\tt shestakova@sfedu.ru}}

\Abstract{Last years a certain attention was attracted to the statement that Hamiltonian formulations of General Relativity, in which different parametrizations of gravitational variables were used, may not be related by a canonical transformation. The example was given by the Hamiltonian formulation of Dirac and that of Arnowitt -- Deser -- Misner. It might witness for non-equivalence of these formulations and the original (Lagrangian) formulation of General Relativity. The problem is believed to be of importance since many authors make use of various representations of gravitational field as a starting point in searching a way to reconcile the theory of gravity with quantum principles. It can be shown that the mentioned above conclusion about non-equivalence of different Hamiltonian formulations is based on the consideration of canonical transformations in phase space of physical degrees of freedom only, while the transformations also involve gauge degrees of freedom. We shall give a clear proof that Hamiltonian formulations corresponding to different parametrizations of gravitational variables are related by canonical transformations in extended phase space embracing gauge degrees of freedom on an equal footing with physical ones. It will be demonstrated for the full gravitational theory in a wide enough class of parametrizations and gauge conditions.}

\section{Introduction}
It is generally believed that the problem of formulating Hamiltonian dynamics for systems with constraints has been solved by Dirac in his seminal papers \cite{Dirac1,Dirac2}. Further, Dirac himself applied his approach to gravitational field \cite{Dirac3}. It is also believed that Dirac generalized Hamiltonian dynamics is equivalent to Lagrangian dynamics of original theory. However, there exist a problem related to the fact that historically, while constructing Hamiltonian dynamics for gravity different authors used various parametrizations of gravitational variables. While Dirac dealt with original variables, which are components of metric tensor \cite{Dirac3}, the most known parametrization is probably that of Arnowitt -- Deser -- Misner (ADM) \cite{ADM}, who expressed $g_{00}$, $g_{0\mu}$ through the lapse and shift functions $N$, $N^i$. Recently it has been shown in \cite{KK} that components of metric tensor and the ADM variables are not related by a canonical transformation. In other words, it implies that the Dirac Hamiltonian formulation for gravitation and the ADM one are not equivalent, though it is supposed that each of them is equivalent to the Einstein (Lagrangian) formulation. There appears the contradiction that witnesses about the incompleteness of the theoretical foundation.

The origin of the contradiction is that the transformation from components of metric tensor to the ADM variables touches gauge degrees of freedom that are not canonical from the viewpoint of the Dirac approach, i.e. it would be impossible to express their velocities in terms of the conjugate momenta:
\begin{equation}
\label{ADM-tr}
g_{00}=\gamma_{ij}N^i N^j-N^2,\qquad
g_{0i}=\gamma_{ij}N^j,\qquad
g_{ij}=\gamma_{ij}.
\end{equation}
To pose the question, if the transformation (\ref{ADM-tr}) is canonical, one should formally extend the original phase space including into it gauge degrees of freedom and their momenta. In particular, one should check if the Poisson brackets among all gravitational variables and their momenta maintain their form after the transformation (\ref{ADM-tr}). To prove non-canonicity of (\ref{ADM-tr}) it is enough to check that some of the Poisson brackets relations are broken. For the sake of generality, let us consider the inverse ADM-like transformation
\begin{equation}
\label{ADMl-tr}
N_{\mu}=V_{\mu}(g_{0\nu},\, g_{ij}),\qquad
\gamma_{ij}=g_{ij}.
\end{equation}
Here $V_{\mu}$ are some functions of components of metric tensor (but $N_{\mu}$ ought not to form 4-vector). A feature of this transformation is that space components of metric tensor remain unchanged, and so do their conjugate momenta: $\Pi^{ij}=p^{ij}$. Then
\begin{equation}
\label{ADM-PB}
\left.\{N_{\mu},\,\Pi^{ij}\}\right|_{g_{\nu\lambda},p^{\rho\sigma}}=\frac{\partial V_{\mu}}{\partial g_{ij}}.
\end{equation}
It is equal to zero if only the functions $V_{\mu}$ do not depend on $g_{ij}$. This is quite a trivial case when old gauge variables are expressed through some new gauge variables only, and the ADM transformation (\ref{ADM-tr}) does not belong to this class.

From the viewpoint of the Lagrangian formalism, different parametrizations are admissible and corresponding Lagrange formulations are equivalent. Equations of motions and constraints can be obtained in the Lagrangian formalism by the unique variational procedure. In the generalized Hamiltonian dynamics constraints have a special status, different from that of Hamiltonian motion equation. Reparametrizations like (\ref{ADM-tr}), (\ref{ADMl-tr}) violate the algebra of constraints and, therefore, lead to non-equivalence of Hamiltonian formulations. In order to construct a Hamiltonian dynamics that would be fully equivalent to the Lagrangian dynamics, one should not just extend phase space of canonical variables including formally gauge degrees of freedom in it, but also introduce missing velocities into the Lagrangian by means of special (differential) gauge conditions that actually extends the phase space. This way was outlined in our papers \cite{SSV1,SSV2}, and now we are able to demonstrate for full gravitational theory that different parametrizations in a wide enough class are related by canonical transformations in extended phase space.

\section{The effective action for the full gravitational theory}
We shall consider the effective action including gauge and ghost sectors as it appears in the path integral approach to gauge field theories,
\begin{equation}
\label{full-act}
S=\int d^4 x\left({\cal L}_{(grav)}+{\cal L}_{(gauge)}+{\cal L}_{(ghost)}\right)
\end{equation}

As any Lagrangian is determined up to a total derivative, the gravitational Lagrangian density ${\cal L}_{(grav)}$ can be modified in such a way for the primary constraints to take the form \cite{Dirac3}
\begin{equation}
\label{primary}
\pi^{\mu}=0,
\end{equation}
where $\pi^{\mu}$ are the momenta conjugate to gauge variables $g_{0\mu}$. This change of the Lagrangian density does not affect the equation of  motion.

We shall use a gauge condition in a general form, $f^{\mu }(g_{\nu\lambda})=0$. The differential form of this gauge condition introduces the missing velocities and actually extends phase space,
\begin{equation}
\label{diff-g}
\frac{d}{dt}f^{\mu}(g_{\nu\lambda})=0,\qquad
\frac{\partial f^{\mu}}{\partial g_{00}}\dot g_{00}
 +2\frac{\partial f^{\mu}}{\partial g_{0i}}\dot g_{0i}
 +\frac{\partial f^{\mu}}{\partial g_{ij}}\dot g_{ij}=0.
\end{equation}
Then,
\begin{equation}
\label{L_gauge}
{\cal L}_{(gauge)}=\lambda_{\mu}\left(\frac{\partial f^{\mu}}{\partial g_{00}}\dot g_{00}
  +2\frac{\partial f^{\mu}}{\partial g_{0i}}\dot g_{0i}
  +\frac{\partial f^{\mu}}{\partial g_{ij}}\dot g_{ij}\right).
\end{equation}
Taking into account the gauge transformations,
\begin{equation}
\label{g-tr}
\delta g_{\mu\nu}=\partial_{\lambda}g_{\mu\nu}\theta^{\lambda}
  +g_{\mu\lambda}\partial_{\nu}\theta^{\lambda}
  +g_{\nu\lambda}\partial_{\mu}\theta^{\lambda},
\end{equation}
one can write the ghost sector:
\begin{equation}
\label{L_ghost}
{\cal L}_{(ghost)}=\bar\theta_{\mu}\frac{d}{dt}
  \left[\frac{\partial f^{\mu}}{\partial g_{\nu\lambda}}
   \left(\partial_{\rho}g_{\nu\lambda}\theta^{\rho}
     +g_{\lambda\rho}\partial_{\nu}\theta^{\rho}
     +g_{\nu\rho}\partial_{\lambda}\theta^{\rho}\right)\right].
\end{equation}

It is convenient to write down the action (\ref{full-act}), (\ref{L_gauge}), (\ref{L_ghost}) in the form
\begin{eqnarray}
S&=&\int d^4 x\left[{\cal L}_{(grav)}+\Lambda_{\mu}
  \left(\frac{\partial f^{\mu}}{\partial g_{00}}\dot g_{00}
   +2\frac{\partial f^{\mu}}{\partial g_{0i}}\dot g_{0i}
   +\frac{\partial f^{\mu}}{\partial g_{ij}}\dot g_{ij}\right)
  -\dot{\bar{\theta_{\mu}}}\left(\frac{\partial f^{\mu}}{\partial g_{00}}
    \left(\partial_i g_{00}\theta^i+2g_{0\rho}\dot\theta^{\rho}\right)\right.\right.\nonumber\\
\label{action1}
&+&\left.\left.2\frac{\partial f^{\mu}}{\partial g_{0i}}
  \left(\partial_j g_{0i}\theta^{j}+g_{0\rho}\partial_i\theta^{\rho}+g_{i\rho}\dot\theta^{\rho}\right)
 +\frac{\partial f^{\mu}}{\partial g_{ij}}
  \left(\partial_k g_{ij}\theta^k+g_{i\rho}\partial_j\theta^{\rho}+g_{j\rho}\partial_i\theta^{\rho}\right)\right)\right].
\end{eqnarray}
Here $\Lambda_{\mu}=\lambda_{\mu}-\dot{\bar{\theta_{\mu}}}\theta^0$. One can see that the generalized velocities enter into the bracket multiplied by $\Lambda_{\mu}$, in addition to the gravitational part ${\cal L}_{(grav)}$. This very circumstance will ensure the canonicity of the transformation to new variables.

\section{The transformation to new variables}
Our goal now is to introduce new variables by
\begin{equation}
\label{new-var}
g_{0\mu}=v_{\mu}\left(N_{\nu},g_{ij}\right).
\end{equation}
This transformation concerns only $g_{0\mu}$ metric components. In a particular case, $N_{\mu}$ may be expressed through the lapse and shift functions introduced by ADM, $v_{\mu}$ are invertible functions, so that
\begin{equation}
\label{inv-tr}
N_{\mu}=V_{\mu}\left(g_{0\nu},g_{ij}\right).
\end{equation}
After the transformation (\ref{new-var}) the action will read
\begin{eqnarray}
S&=&\int d^4 x\left[{\cal L'}_{(grav)}
   +\Lambda_{\mu}\left(\frac{\partial f^{\mu }}{\partial g_{00}}\;
     \frac{\partial v_0}{\partial N_{\lambda }}\;\dot N_{\lambda}
    +\frac{\partial f^{\mu}}{\partial g_{00}}\;
     \frac{\partial v_0}{\partial g_{ij}}\;\dot g_{ij}
    +2\;\frac{\partial f^{\mu}}{\partial g_{0i}}\;
     \frac{\partial v_i}{\partial N_{\lambda}}\;\dot N_{\lambda}+\right.\right.\nonumber\\
&+&\left.2\;\frac{\partial f^{\mu}}{\partial g_{0k}}\;
     \frac{\partial v_k}{\partial g_{ij}}\;\dot g_{ij}
    +\frac{\partial f^{\mu}}{\partial g_{ij}}\;\dot g_{ij}\right)
   -\dot{\bar{\theta_{\mu}}}\left(\frac{\partial f^{\mu }}{\partial g_{00}}\;
     \frac{\partial v_0}{\partial N_{\lambda}}\;\partial_i N_{\lambda}\theta^i
    +\frac{\partial f^{\mu}}{\partial g_{00}}\;
     \frac{\partial v_0}{\partial g_{ij}}\;\partial_k g_{ij}\theta^k+\right.\nonumber\\
&+&2\;\frac{\partial f^{\mu}}{\partial g_{00}}\;v_{\sigma }(N_{\tau },\,g_{ij})\;\dot{\theta}^{\sigma}
    +2\;\frac{\partial f^{\mu}}{\partial g_{0i}}\;
     \frac{\partial v_i}{\partial N_{\lambda}}\;\partial_j N_{\lambda}\theta^j
    +2\;\frac{\partial f^{\mu}}{\partial g_{0i}}\;
     \frac{\partial v_i}{\partial g_{ij}}\;\partial_k g_{ij}\theta^k\nonumber\\
&+&2\;\frac{\partial f^{\mu}}{\partial g_{0i}}\left[v_{\rho}(N_{\tau},\;g_{ij})\partial_i\theta^{\rho}
    +v_i(N_{\sigma},\;g_{ij})\dot\theta^0+g_{ij}\dot\theta^j\right] +\nonumber\\
\label{action2}
&+&\left.\left.\frac{\partial f^{\mu}}{\partial g_{ij}}\left[\partial_k g_{ij}\theta^k
    +v_i(N_{\sigma},\;g_{ij})\partial_j\theta^0+g_{ik}\partial_j\theta^k
    +v_j(N_{\sigma},\;g_{ij})\partial_i\theta^0+g_{jk}\partial_i\theta^k\right]\right)\right]
\end{eqnarray}

We can write down the ``old'' momenta,
\begin{equation}
\label{old-mom}
\pi^{ij}=\frac{\partial{\cal L}_{(grav)}}{\partial\dot g_{ij}}
 +\Lambda_{\mu}\frac{\partial f^{\mu}}{\partial g_{ij}};\qquad
\pi^0=\frac{\partial{\cal L}_{(grav)}}{\partial\dot g_{00}}
 +\Lambda_{\mu}\frac{\partial f^{\mu}}{\partial g_{00}};\qquad
\pi^i=\frac{\partial{\cal L}_{(grav)}}{\partial\dot g_{0i}}
 +2\Lambda_{\mu}\frac{\partial f^{\mu}}{\partial g_{0i}},
\end{equation}
and the ``new''momenta are:
\begin{eqnarray}
\Pi^{ij}&=&\frac{\partial{\cal L'}_{(grav)}}{\partial\dot g_{ij}}
 +\Lambda_{\mu}\left(\frac{\partial f^{\mu}}{\partial g_{00}}\;
   \frac{\partial v_0}{\partial g_{ij}}
 +2\;\frac{\partial f^{\mu}}{\partial g_{0k}}\;
   \frac{\partial v_k}{\partial g_{ij}}
 +\frac{\partial f^{\mu}}{\partial g_{ij}}\right);\nonumber\\
\label{new-mom}
\Pi^0&=&\frac{\partial{\cal L'}_{(grav)}}{\partial\dot N_0}
 +\Lambda_{\mu}\left(\frac{\partial f^{\mu}}{\partial g_{00}}\;
   \frac{\partial v_0}{\partial N_0}
 +2\;\frac{\partial f^{\mu}}{\partial g_{0i}}\;
   \frac{\partial v_i}{\partial N_0}\right);\\
\Pi^i&=&\frac{\partial{\cal L'}_{(grav)}}{\partial\dot N_i}
 +\Lambda_{\mu}\left(\frac{\partial f^{\mu}}{\partial g_{00}}\;
  \frac{\partial v_0}{\partial N_i}
 +2\;\frac{\partial f^{\mu}}{\partial g_{0j}}\;
   \frac{\partial v_j}{\partial N_i}\right).\nonumber
\end{eqnarray}

The relations between the ``old'' and ``new'' momenta:
\begin{equation}
\label{relat1}
\Pi^{ij}=\pi^{ij}+\left(\pi^{\mu}
 -\frac{\partial{\cal L}_{(grav)}}{\partial\dot g_{0\mu}}\right)
   \frac{\partial v_{\mu}}{\partial g_{ij}};\qquad
\Pi^{\mu}=\frac{\partial{\cal L'}_{(grav)}}{\partial\dot N_{\mu}}
 +\left(\pi^{\nu}-\frac{\partial{\cal L}_{(grav)}}{\partial\dot g_{0\nu}}\right)
   \frac{\partial v_{\nu}}{\partial N_{\mu}}.
\end{equation}
It is easy to check that the momenta conjugate to ghosts remain unchanged, $\tilde{\cal P}^{\mu}={\cal P}^{\mu}$, 
$\tilde{\bar{\cal P}}_{\mu}=\bar{\cal P}_{\mu}$.

As it has been already mentioned (see (\ref{primary})), the gravitational part of the Lagrangian density can be modified so that
\begin{equation}
\label{deriv}
\frac{\partial{\cal L}_{(grav)}}{\partial\dot g_{0\mu}}=0,\qquad
\frac{\partial{\cal L'}_{(grav)}}{\partial\dot N_{\mu}}=0.
\end{equation}
Then, the relations (\ref{relat1}) would become simpler and take the form
\begin{equation}
\label{relat2}
\Pi^{ij}=\pi^{ij}+\pi^{\mu}\frac{\partial v_{\mu}}{\partial g_{ij}};\qquad
\Pi^{\mu}=\pi^{\nu}\frac{\partial v_{\nu}}{\partial N_{\mu}}.
\end{equation}

It is easy to demonstrate that the transformations (\ref{inv-tr}), (\ref{relat2}) are canonical in extended phase space. The generating function depends on new coordinates and old momenta \cite{LL},
\begin{equation}
\label{gen-fun}
\Phi\left(N_{\mu },\;g_{ij},\;\theta^{\mu},\;\bar{\theta}_{\mu},\;
   \pi^{\mu},\;\pi^{ij},\;\bar{\cal P}_{\mu},\;{\cal P}^{\mu}\right)
 =-\pi^{\mu}v_{\mu}(N_{\nu},\;g_{ij})-\pi^{ij}g_{ij}
  -\bar{\cal P}_{\mu}\theta^{\mu}-\bar{\theta}_{\mu}{\cal P}^{\mu }.
\end{equation}

Then the following relations take place
\begin{equation}
\label{can-rel1}
g_{0\mu}=-\frac{\partial\Phi}{\partial\pi^{\mu}};\qquad
g_{ij}=-\frac{\partial\Phi}{\partial\pi^{ij}};\qquad
\theta^{\mu}=-\frac{\partial\Phi}{\partial\bar{\cal P}\vphantom{\sqrt N}_{\mu}};\qquad
\bar{\theta}_{\mu}=-\frac{\partial\Phi}{\partial{\cal P}^{\mu }};
\end{equation}
\begin{equation}
\label{can-rel2}
\Pi^{\mu}=-\frac{\partial\Phi}{\partial N_{\mu}};\qquad
\Pi^{ij}=-\frac{\partial\Phi}{\partial g_{ij}};\qquad
\bar{\cal P}_{\mu}=-\frac{\partial\Phi}{\partial\theta^{\mu}};\qquad
{\cal P}^{\mu}=-\frac{\partial\Phi}{\partial\bar{\theta}\vphantom{\sqrt N}_{\mu}},
\end{equation}
that give exactly the transformations
\begin{equation}
\label{c-rel1}
g_{0\mu}=v_{\mu}(N_{\nu},\;g_{ij});\qquad
g_{ij}=g_{ij};\qquad\qquad\qquad\qquad
\theta^{\mu}=\theta^{\mu};\qquad
\bar{\theta}_{\mu}=\bar{\theta}_{\mu };
\end{equation}
\begin{equation}
\label{c-rel2}
\Pi^{\mu }=\pi^{\nu}\frac{\partial v_{\nu}}{\partial N_{\mu}};\qquad\qquad
\Pi^{ij}=\pi^{ij}+\pi^{\mu}\frac{\partial v_{\mu}}{\partial g_{ij}};\qquad\quad
\bar{\cal P}_{\mu}=\bar{\cal P}_{\mu};\qquad
{\cal P}^{\mu}={\cal P}^{\mu}.
\end{equation}

We can now check if the Poisson brackets maintain their form. For example, we can recalculate the bracket (\ref{ADM-PB}) to see that it will be zero in our extended phase space formalism.
\begin{eqnarray}
\left.\{N_{\mu},\,\Pi^{ij}\}\right|_{g_{\nu\lambda},p^{\rho\sigma}}
 &=&\frac{\partial N_{\mu}}{\partial g_{0\rho}}\frac{\partial \Pi^{ij}}{\partial\pi^{\rho}}
  +\frac{\partial N_{\mu}}{\partial g_{kl}}\frac{\partial\Pi^{ij}}{\partial\pi^{kl}}
  =\left\{V_{\mu}(g_{0\nu},g_{kl}),\;\pi^{ij}+\pi^{\lambda}
    \frac{\partial v_{\lambda}}{\partial g_{ij}}\right\}\nonumber\\
\label{EPS-PB}
 &=&\frac{\partial V_{\mu}}{\partial g_{0\rho}}
    \frac{\partial v_{\lambda}}{\partial g_{ij}}\delta_{\rho}^{\lambda}
  +\frac{\partial V_{\mu}}{\partial g_{kl}}\frac12\left(\delta_k^i\delta_l^j+\delta_l^j\delta_k^i\right)
=\frac{\partial V_{\mu}}{\partial g_{0\lambda}}\frac{\partial v_{\lambda}}{\partial g_{ij}}
  +\frac{\partial V_{\mu}}{\partial g_{ij}}=0.
\end{eqnarray}
(The last equality in (\ref{EPS-PB}) is due to the rules of implicit differentiation.)

\section{Conclusions}
As we have seen, the extension of phase space by introducing the missing velocities changes the relations between the ``old'' and ``new'' momenta (see (\ref{relat2})). As a consequence, the transformations (\ref{new-var}) in a wide enough class of parametrizations are canonical. In that way, we consider extended phase space not just as an auxiliary construction which enables one to compensate residual degrees of freedom and regularize a path integral, as it was in the Batalin -- Fradkin -- Vilkovisky approach \cite{BFV1,BFV2,BFV3}, but  rather as a structure that ensures equivalence of Hamiltonian dynamics for a constrained system and Lagrangian formulation of the original theory. An important role is played by gauge degrees of freedom, which cannot be excluded from consideration.

There exist another problem how to construct a generator of gauge transformation in phase space, since the Dirac prescription, according to which the generator is given by a linear combination of constraints, cannot guarantee correct transformations for gauge degrees of freedom. We expect that the extended phase space approach would help to find an unambiguous solution to the problem.

In our previous papers \cite{SSV1,SSV2} it has been demonstrated that it is impossible to construct a mathematically consistent quantum theory of gravity without taking into account the role of gauge degrees of freedom in description of quantum gravitational phenomena from the point of view of different observers, In a certain sense, the extended phase space approach follows the spirit of General Relativity and Quantum Theory since all observers are treated as equitable, though different observers can see various, but complementary pictures.

\end{document}